 \documentclass[final,3p,times,twocolumn]{elsarticle}

\usepackage{graphicx, booktabs, enumitem}
\usepackage{epsfig}
\usepackage{dcolumn}
\usepackage[flushleft]{threeparttable}

\usepackage{amssymb}
\usepackage{color}
\usepackage{soul} 

\biboptions{sort,compress}
\journal{Physics Letters B}

\begin{document}

\newcommand{\gray}{$\gamma$-ray}
\newcommand{\jpm}{J$ ^{\pi }$=1$ ^{-} $}
\newcommand{\jppi}{J$^{\pi }$=1$^{+} $}
\newcommand{\jpt}{J$^{\pi } $=2$ ^{+} $}
\newcommand{\Rskin}{R_{\hspace{0.5pt}\rm skin}}
\newcommand{\alphaD}{\alpha_{{}_{\hspace{-0.5pt}D}}}

\begin{frontmatter}



\title{Pygmy and core polarization dipole modes in $^{206}$Pb:\\
connecting nuclear structure to stellar nucleosynthesis}


\author[LLNL,Duke]{A.P. Tonchev\corref{cor1}}
\ead{tonchev2@llnl.gov}
\address[LLNL]{Nuclear and Chemical Sciences Division, Lawrence Livermore National Laboratory, 
Livermore, California 94550, USA}
\address[Duke]{Duke University, Department of Physics, Box 90308, Durham, North Carolina 27708-0308, USA}

\author[giess]{N. Tsoneva}
\address[giess]{Institut f{\"u}r Theoretische Physik, Universit\"{a}t Gie$\beta$en, Heinrich-Buff-Ring 16, D-35392 Gie$\beta$en, Germany}

\author[India,Duke,TUNL]{C. Bhatia}
\address[India]{Society for Applied Microwave Electronics Engineering and Research, Indian Institute of Technology, Mumbai Campus, India}
\address[TUNL]{Triangle Universities Nuclear Laboratory, Durham, North Carolina 27708-0308, USA}

\author[BAPL,LANL]{C.W. Arnold}
\address[BAPL]{Bettis Atomic Power Laboratory, West Mifflin, PA 15122}
\address[LANL]{Chemistry Division, Los Alamos National Laboratory, Los Alamos, New Mexico 87545, USA}

\author[UB]{S.\,Goriely}
\address[UB]{Institut d'Astronomie et d'Astrophysique, Universit\'e Libre de Bruxelles, Campus de la Plaine, CP-226, 1050 Brussels, Belgium}

\author[UNC,TUNL]{S.L. Hammond}
\address[UNC]{Department of Physics and Astronomy, University of North Carolina at Chapel Hill, Chapel Hill, North Carolina 27599-3255, USA}

\author[UNC,TUNL]{J.H. Kelley}

\author[NSCL]{E. Kwan}
\address[NSCL]{National Superconducting Cyclotron Laboratory, East Lansing, Michigan 48824, USA}

\author[giess]{H. Lenske}

\author[FSU]{J. Piekarewicz}
\address[FSU]{Department of Physics, Florida State University, Tallahassee, Florida 32306-4350, USA}

\author[KC]{R.\,Raut}
\address[KC]{UGC-DAE Consortium for Scientific Research, Kolkata Centre LB-8 Sector-III, Bidhannagar, Kolkata 700098, India}

\author[LANL]{G. Rusev}

\author[JAEA]{T. Shizuma}
\address[JAEA]{Quantum Beam Science Center, Japan Atomic Energy Agency, Tokai-mura, Ibaraki 319-1184, Japan}

\author[Duke,TUNL]{W. Tornow}

\cortext[cor1]{Corresponding author}

\begin{abstract}
A high-resolution study of the electromagnetic response of $^{206}$Pb below the neutron separation 
 energy is performed using a ($\vec{\gamma}$,$\gamma'$) experiment at the HI$\vec{\gamma}$S 
 facility. Nuclear resonance fluorescence with 100\% linearly polarized photon beams is used to measure
 spins, parities, branching ratios, and decay widths of excited states in $^{206}$Pb from $4.9$ to $8.1$\,MeV. 
The extracted $\Sigma$$B$(E1)$\uparrow$  and $\Sigma$$B$(M1)$\uparrow$ values 
 for the total electric and magnetic dipole strength below the neutron separation energy are 
 0.9$\pm$0.2\,e$^2$fm$^2$ and 8.3\,$\pm$\,2.0\,$\mu_{N}^2$, respectively. These measurements
 are found to be in very good agreement with the predictions from an energy-density functional (EDF) 
 plus quasiparticle phonon model (QPM). Such a detailed theoretical analysis allows to 
 separate the pygmy dipole resonance from both the tail of the giant dipole resonance and multi-phonon 
 excitations. Combined with earlier photonuclear experiments above the neutron separation energy, one 
 extracts a value for the electric dipole polarizability of $^{206}$Pb of $\alphaD\!=\!122\pm10$\,mb/MeV. 
 When compared to predictions from both the EDF+QPM and accurately calibrated relativistic EDFs, 
 one deduces a range for the neutron-skin thickness of $R_{\rm skin}^{206}\!=\!0.12$-$0.19$\,fm and a
 corresponding range for the slope of the symmetry energy of  $L\!=\!48$-$60$\,MeV. This newly obtained 
 information is also used to estimate the Maxwellian-averaged radiative cross section 
 $^{205}$Pb(n,$\gamma$)$^{206}$Pb at 30\,keV to be $\sigma\!=\!130\!\pm\!25$\,mb. The
 astrophysical impact of this measurement---on both the s-process in stellar nucleosynthesis and on
 the equation of state of neutron-rich matter---is discussed.  
\end{abstract}


\end{frontmatter}


The commissioning of both powerful telescopes and sophisticated 
terrestrial facilities has resulted in a strong synergy between nuclear physics 
and astrophysics. \emph{Where do the chemical elements come from?} and 
\emph{what is the nature of matter at extreme densities?} are a few of the
fundamental questions animating nuclear astrophysics today. However, 
progress in answering these questions relies on our understanding of 
nuclear structure far away from the valley of stability where the neutron-proton 
asymmetry is large and our knowledge is poor. The dynamics of such exotic 
systems is imprinted in the nuclear symmetry energy:
\begin{equation}
 S(\rho) \equiv \frac{1}{2}
 \left(\frac{\partial^{2}{\mathcal E}(\rho,\delta)}{\partial\delta^{2}}\right)_{\!\!\delta=0}
 \hspace{-10pt}\approx \mathcal{E}(\rho,\delta\!=\!1)-\mathcal{E}(\rho,\delta\!=\!0).
 \label{SymmE}
\end{equation}
Here ${\mathcal E}(\rho,\delta)$ is the energy per nucleon as a function of the 
total baryon density $\rho$ and  the neutron-proton asymmetry 
$\delta\!\equiv\!(N\!-\!Z)/A$. Note that to a very good approximation
the symmetry energy represents the energy cost of turning symmetric nuclear matter 
($\delta\!=\!0$) into pure neutron matter ($\delta\!=\!1$). In particular, the density 
dependence of the symmetry energy around nuclear saturation density ($\rho_{0}$) 
is encoded in a few bulk parameters\,\cite{Pie09}:
\begin{equation}
 S(\rho) = J + Lx + \frac{1}{2}K_{\rm sym}x^{2}+\ldots   \hspace{10pt}{\rm with}
 \hspace{3pt} x\equiv\frac{\rho - \rho_{0}}{3\rho_{0}}.
\label{JLK}
\end{equation}
An observable that is particularly sensitive to the density dependence of the 
symmetry energy is the neutron-skin thickness of a heavy nucleus, defined 
as the difference between the neutron and proton rms radii 
$\Rskin\!=\!R_{n}\!-\!R_{p}$. In particular, the neutron skin of ${}^{208}$Pb 
is strongly correlated to the \emph{slope} of the symmetry energy 
$L$\,\cite{Bro00,Cen09}. The Lead Radius Experiment (PREX) has 
provided the first model independent evidence in favor of a neutron-rich skin 
in ${}^{208}$Pb\,\cite{Abr12,Hor12}. Unfortunately, unforeseen technical issues 
compromised the statistical accuracy of the measurement. And whereas a 
follow-up experiment (PREX-II) is envisioned to reach the original sensitivity, 
a complementary observable has been identified that is also strongly correlated 
to the slope of the symmetry energy: the electric dipole polarizability 
$\alphaD$\,\cite{Rei10,Tam11}. 

The Pygmy Dipole Resonance (PDR)---the emergence of low-energy dipole 
strength with neutron excess---has motivated a great deal of experimental and 
theoretical effort\,\cite{Sav13}. Observed as a concentration of electromagnetic 
strength overlapping the low-energy tail of the Giant Dipole Resonance (GDR), 
the PDR has been identified in a broad range of neutron-rich systems, ranging 
from light\,\cite{Ina11,Der14}, to transitional\,\cite{Rus09,Mas14}, up to heavy 
nuclei\,\cite{Rye02,End03,End04,Tam11} far away from stability. 
In this regard the PDR---and perhaps higher-multipolarity 
modes \cite{Tso11,Pel15,Spi16}---may provide useful constraints on such fundamental properties as the neutron-skin 
thickness of medium to heavy nuclei\,\cite{Pie06,Tam11,Pie11}, the nuclear symmetry 
energy\,\cite{Ina11,Maz15}, and the properties of neutron stars\,\cite{Hor01}.

The heavy $^{206}$Pb nucleus containing 42 excess neutrons is expected 
to have a robust neutron skin, and as a consequence, should exhibit an appreciable amount of low-energy dipole strength.
“A systematic study of the nuclear dipole strength in the lead isotopes, including $^{206}$Pb, is presented in Ref. [14] for excitation energies up to 6.5 MeV.”
 At excitation energies below the neutron separation energy ($S_n\!=\!8.087$\,MeV)
the PDR coexists with a variety of modes, such as the tail of the GDR, magnetic dipole transitions,
and multi-phonon excitations\,\cite{Hey10,Tso04,Tso08,Tso15}. So far, isolating the low-energy
photoabsorption spectra from these other contributions has proved elusive\,\cite{Rus13,Kri15,Ina15}.
As a result and despite significant experimental and theoretical effort, see e.g. \cite{Ina15}, a direct model-independent determination of the neutron-skin thickness from the PDR is not yet possible\,\cite{Sav13}.
However, rather than concentrating on the fraction of the PDR that contributes to the cross section, a more 
robust observable is the ``inverse-squared" energy weighted sum ($\sigma_{-2}$). That is\,\cite{Boh81},
\begin{equation}
 \alphaD =  \frac{1}{2\pi^{2}\alpha} \int_{0}^{\infty} 
   \frac{\sigma_{\gamma}(E)}{E^{2}}\,dE
 =\frac{\sigma_{-2}}{2\pi^{2}\alpha} = 6.942\,\sigma_{-2},
 \label{EDP}
\end{equation}  
where $\sigma_{-2}$ is directly proportional to the electric dipole polarizability 
$\alphaD$ (both given in mb/MeV) and $\alpha$ is the fine-structure constant. 

As far as the impact of the present measurement on stellar nucleosynthesis is concerned, the PDR in 
$^{206}$Pb might affect the $^{205}$Pb radiative neutron capture cross section, a reaction 
of relevance to the destruction of $^{205}$Pb during the s-process \cite{Rau13, Tso15}. 
The now extinct radionuclide $^{205}$Pb with a half-life of 15.3\,Myr may be of significant cosmochemical interest due to its pure s-process nature\,\cite{Yokoi85}. 
Further, it could provide key information on the formation 
of the solar system, particularly on the time span between the last s-process nucleosynthetic 
events that have modified the composition of the solar nebula and the formation of the solid 
bodies in the solar system. The presence of $^{205}$Pb in the early solar system was 
demonstrated recently and it was suggested that the $^{205}$Pb-$^{205}$Tl pair is well 
suited for chronological studies, complementing information provided by other extinct 
short-lived radionuclides \cite{Nielsen06,Baker10}. The abundance of $^{205}$Pb in the early 
solar system inferred from carbonaceous chondrites data can also be used to assess whether 
asymptotic giant branch (AGB) stars and massive Wolf-Rayet stars are the most likely sites of 
the s-process\,\cite{Wasserburg06,Arnould97}. However, this investigation remains sensitive 
to the amount of freshly produced s-process material, including the survival of $^{205}$Pb 
in AGB stars\,\cite{Mowlavi98} through both neutron capture and weak interaction processes. 
Due to its astrophysical importance, the Maxwellian-averaged capture cross section on $^{205}$Pb 
needs to be known accurately, and if not measurable directly, at least it should be experimentally constrained.

\begin{figure}
\centering
\includegraphics[width=0.45\textwidth, angle=0]{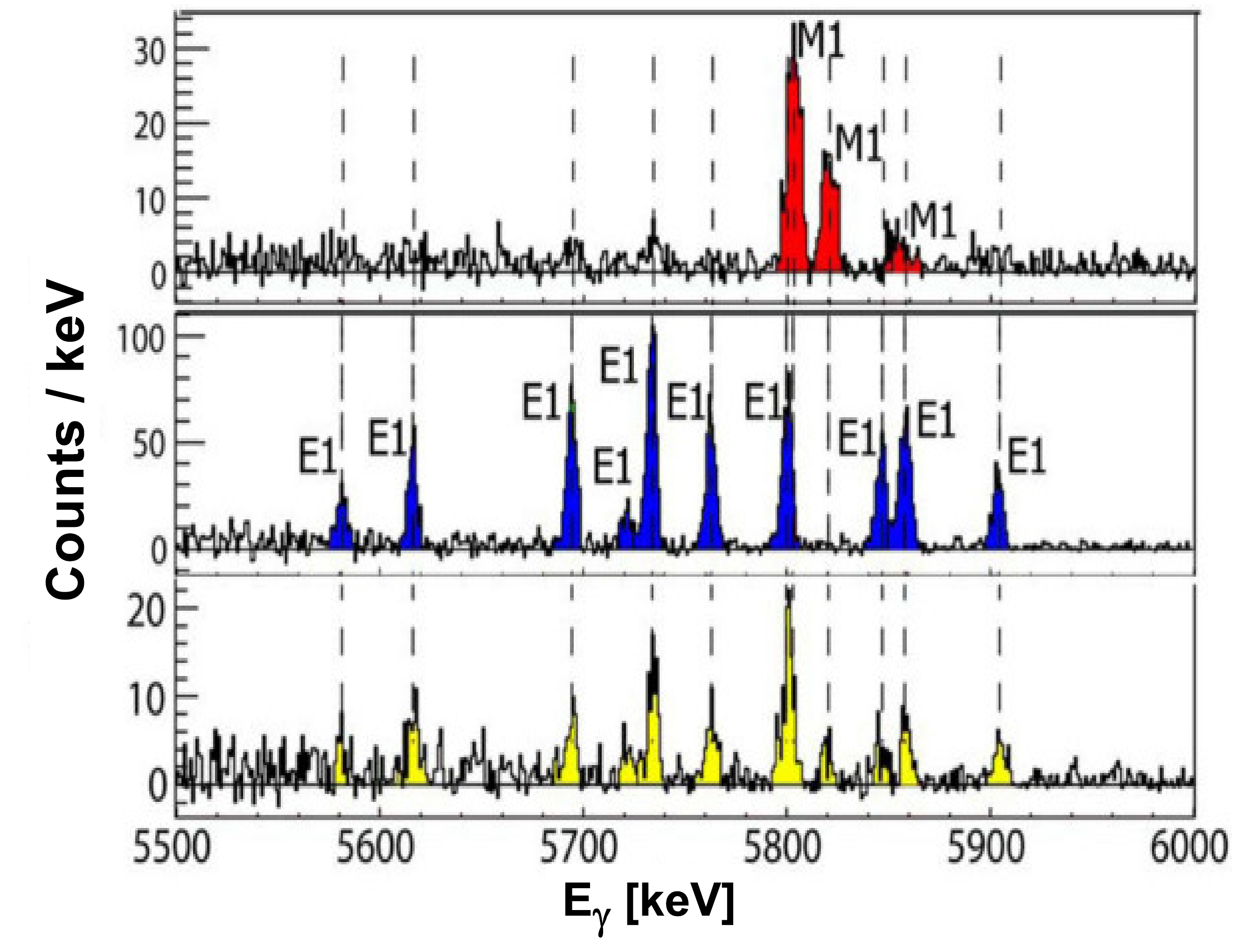}
\caption{(Color online) $\gamma$-ray spectra generated by a photon beam striking a $^{206}$Pb target with a centroid  energy of 5.750\,MeV. The energy spread of the beam is a Gaussian with FWHM = 203 $\pm$ 4 keV. The top spectrum indicates photon scattering ($M$1 transitions) in the polarization plane of the incident beam. The center spectrum indicates photon scattering ($E$1 transitions) perpendicular to the polarization plane. Finally, the bottom spectrum indicates photon scattering at 135$^\circ$  distinguishing $M$1 from $E$2 radiations.}
\label{fig1}
\end{figure}

In the present study we were able to decompose the multipolarity and the decay pattern of the 
PDR and the GDR in $^{206}$Pb below the neutron separation energy. We studied the structure 
of $^{206}$Pb using monoenergetic and 100\% linearly polarized photon beams from the 
High-Intensity Gamma-ray Source (HI$\vec{\gamma}$S) facility\,\cite{Wel08}. Nuclear resonance 
fluorescence (NRF) measurements were performed in the energy range from 4.9 to 8.1 MeV. The 
sample consisted of 4 g of metal powder enriched to 99.3\% in $^{206}$Pb, and contained in a thin-walled and ultra-pure quartz ampule with diameter of 0.9 cm. Possible contributions of resonance states from the SiO$_{2}$ ampule were searched for and found to be negligible, in agreement with literature data  \cite{NNDC}. At each energy, an average photon flux of 10$^7$s$^{-1}$ bombarded the $^{206}$Pb target for approximately four 
hours. Six HPGe detectors (four with 60\% and two with 25\% relative efficiency to a standard 
7.62\,cm\,$\times\,$7.62\,cm NaI detector) were used to measure the  $\gamma$ rays emitted from the NRF process. Two detectors were arranged in the horizontal plane at 90$^\circ$ to the incident beam, 
two were at 90$^\circ$  in the vertical plane, while the remaining two detectors were placed 
at 135$^\circ$ in the horizontal plane. The polarization plane of the beam was the horizontal plane. 
The energy distribution of the photon beams was measured using a 123\% efficiency HPGe detector 
placed a short distance behind the target position. During beam profile measurements, the beam was attenuated by a series of copper absorbers mounted upstream to avoid pileup and long dead times. A portion of the spectra of NRF $\gamma$ rays from a photon beam with 
centroid energy of 5.750 MeV is displayed in Fig.\,1 showing unambiguously a clear distinction between 
$E$1, $M$1, and $E$2 multipolarities\,\cite{Ton10, Rus13}. More detailed information about the NRF 
technique used at the HI$\vec{\gamma}$S facility may be found in 
Refs.\,\cite {Pie02,Sch13,Kri15,God13,Mas14}.  

\begin{figure}[h]
\centering
\includegraphics[width=0.45\textwidth, angle=0]{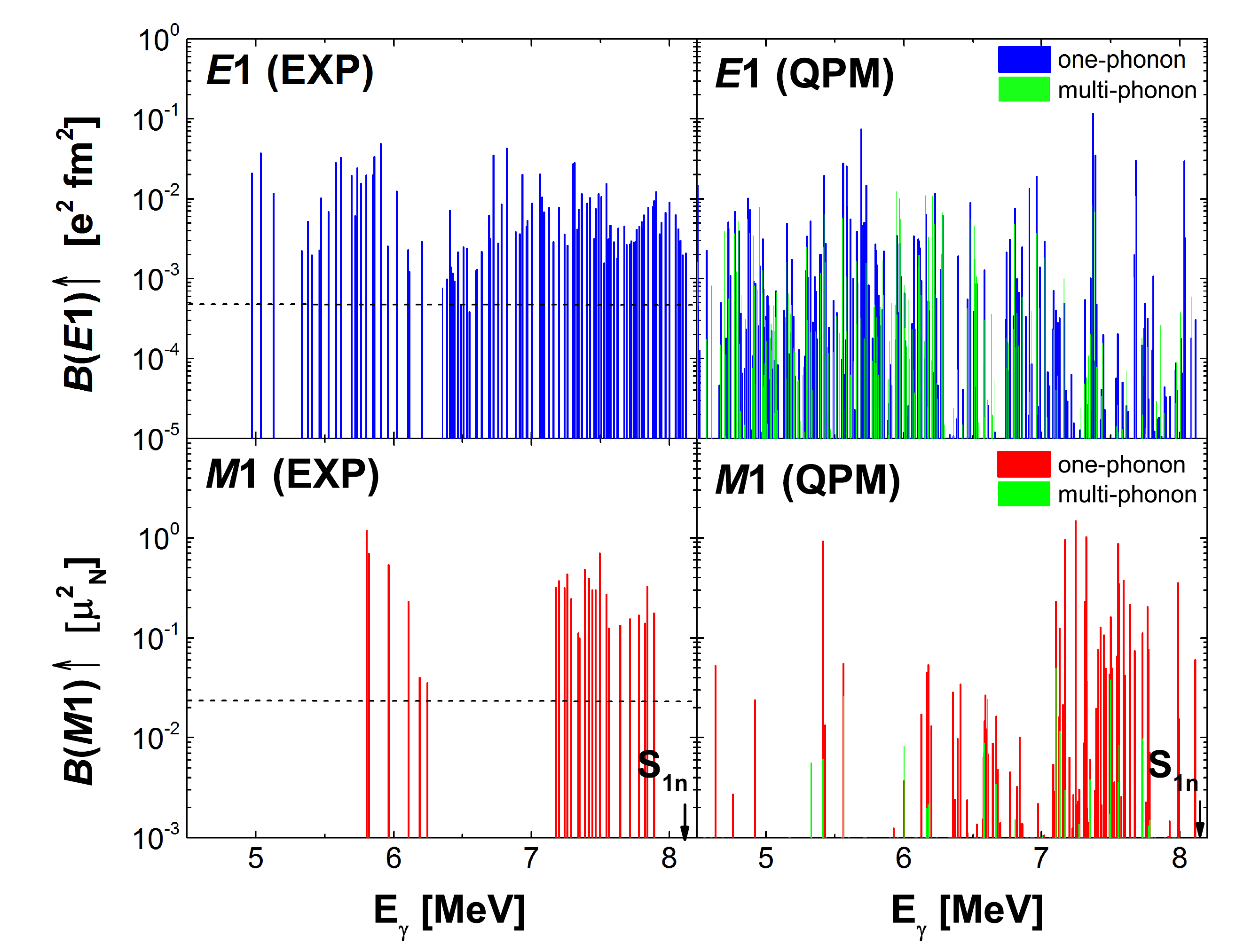}
\caption{(Color online) $B$($E$1) (top panels) and $B$($M$1) (bottom panels) strength 
distribution in $^{206}$Pb for resonantly excited states between 4.9 and 8.1 MeV. Shown is a  comparison between experimental measurements (left) and QPM calculations 
(right) that include a coupling to one- and multi-phonon states. The dashed lines represent 
the detection limit for the $E$1 (top) and $M$1 (bottom) strength. The corresponding 
integrated $B$($E$1) and $B$($M$1) strength is summarized in Table 1.}
\label{fig2}
\end{figure}

The distribution of electric and magnetic dipole states in $^{206}$Pb from 4.9 to 8.1 MeV is 
shown in Fig.\,2. A total of one hundred 1$^-$ states were observed at low energies, with
most of them located at $E_{\gamma}\!\gtrsim\!6.5$\,MeV where the level density is very high. 
Our results for the electric dipole strength are consistent with the presented values in Refs. \cite{End03, Cha80}. However, there are many dipole states missing in the last two references due to the low sensitivity associated with bremsstrahlung beams.
In turn,  twenty-six $M$1 states were identified with the strength localized largely in 
two regions around 6 and 7.5 MeV, assumed to be associated with a spin-flip 
resonance\,\cite{Har01,Rus13,Kri15}. However, it is clear that in this energy region the 
dipole response is predominantly electric in nature. The vast majority of these transitions 
decay directly to the ground state. 

\begin{table}
 \caption{Summary of the $E1$ and $M1$ strengths in $^{206}$Pb.}
    \begin{tabular}{lcr}
       \hline
     Parameter & Present data & EDF+QPM  \\
     \hline
	Energy interval (MeV) & 4.9 - 8.1 & 4.9 - 8.1  \\
	Number of $E$1 states: &  &                  \\
	Within the exp. sensitivity$^{a}$ & 100$^{a}$ & 94  \\
	Total &   & 340  \\
	$\Sigma$$B$(E1) $\uparrow$ (e$^2$fm$^2$) &  0.9 $\pm$ 0.2  &  0.9  \\
     	\\
	Number of $M$1 states: &  &                  \\
	Within the exp. sensitivity$^{b}$ & 26$^{b}$ & 28  \\
	Total &   & 170  \\    
	$\Sigma$$B$(M1) $\uparrow$ ($\mu_{N}^2$) &  8.3 $\pm$ 2.0 &  8.9  \\
 \hline
  \end{tabular}
  \footnotetext(a) {The sensitivity limit for a single $E$1 transition is $\sim\!5\times$10$^{-4}$\,e$^2\,$fm$^2$}\\
 \footnotetext(b) {The sensitivity limit for a single $M$1 transition is $\sim\!\times$10$^{-2}$\,$\mu_{N}^2$}

 \end{table}

Needless to say, a proper description of the rich and complex experimental spectrum depicted 
in Fig.\,\ref{fig2} requires a highly sophisticated theoretical treatment. Thus, displayed
in Fig.\,2 is a detailed comparison of the experimentally observed $E1$ and $M1$ 
strength distributions against theoretical predictions from three-phonon EDF+QPM calculations\cite{Tso04,Tso08}. 
The faithful reproduction of the experimental data is fully consistent 
with earlier EDF+QPM calculations that successfully reproduced the experimental photoneutron 
cross sections---albeit in the GDR region---in $^{206}$Pb, $^{207}$Pb, and $^{208}$Pb\,\cite{Kon12}. 

We emphasize that in contrast to the QPM calculations of Ref.\,\cite{End03}, our EDF+QPM calculations are performed with single-particle energies obtained in a self-consistent manner from our EDF approach based on fully self-consistent Hartree-Fock-Bogoliubov (HFB) calculations. 
In this sense our QPM calculations are considerably more elaborate than those of Ref.\cite{End03}. They account for nuclear ground-state properties like binding energies, root-mean-square radii and the difference between them \cite{Tso08,Tso15}. This is found to be very important for the reliable description of low-energy dipole and higher-multipole pygmy resonances, which are strongly connected to the neutron skin thickness. More details on the comparison between our and other theoretical methods are given in Refs. \cite{Oze14,Rus13}. 

Following our previous EDF+QPM calculations \cite{Ton10,Rus13,Kri15}, the M1 transitions in $^{206}$Pb are calculated with a quenched effective spin-magnetic factor $g_s^{eff}$=0.75$g_s^{bare}$ where the bare spin-magnetic moment g$_s^{bare}$ has been adopted. 
Similar to our findings in the $N\!=\!50,82$ isotones\,\cite{Ton10,Rus13} and in $^{52}$Cr\,\cite{Kri15}, the M1 strength below the neutron separation energy is dominated by spin-flip excitations. A small admixture of about 7$\%$ from orbital contributions is obtained. 

\begin{figure}
 \centering
 \includegraphics[width=0.48\textwidth, angle=0]{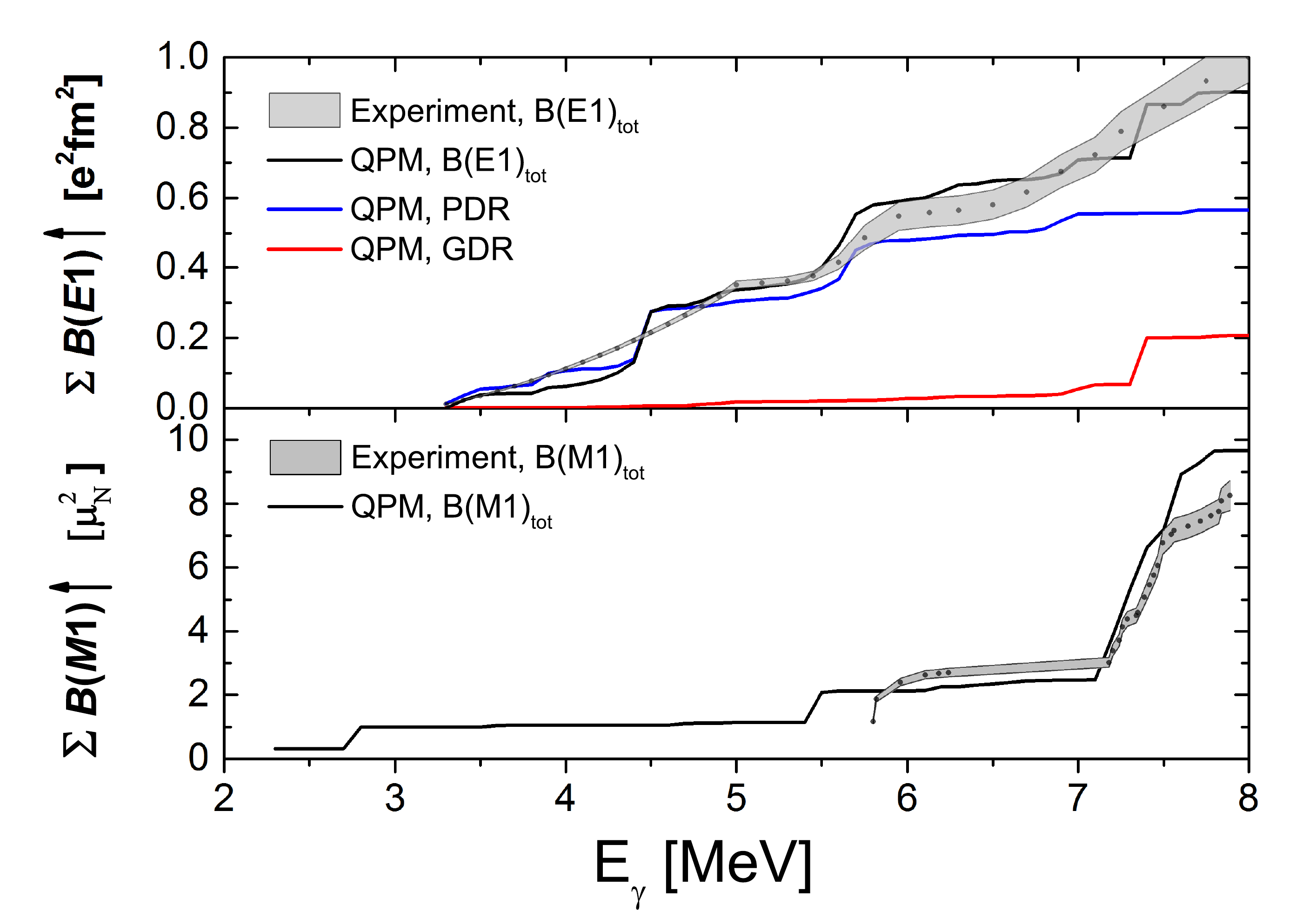}
 \caption{(Color online) Cumulative $B$($E$1) (top) and $B$($M$1) (bottom) strength in 
 $^{206}$Pb obtained from integrating the corresponding distribution of strength up to an
 energy $E_{\gamma}\!\leq\!S_{1n}\!=\!8.087\,{\rm MeV}$.}
  \label{fig3}
\end{figure}

The calculated EDF+QRPA 1$^-$ excitations in the $E_\gamma$ $\leq$ 6.8 MeV region in $^{206}$Pb 
are dominated by almost pure neutron, weakly bound two-quasiparticle configurations which may be 
associated with neutron skin vibrations. The predicted structure of the states strongly resembles our 
previous findings on PDR systematics obtained in a large 
variety of atomic nuclei \cite{Sav13,Tso04,Tso08,Ton10,Rus13,Sch13,Kri15}.
Similar conclusions can be also drawn by means of detailed studies of transition densities, indicating a skin mode with dominantly neutron oscillations at the nuclear surface as is characteristic of the PDR.
With the increase of the excitation energy ($E_\gamma$ $\geq$ 6.8 MeV) the isovector contribution 
to the E1 strength also increases as indicated by the out-of-phase relations of the proton and neutron contributions to the QRPA states \cite{Tso08,Sch13}. 
Moreover, the EDF+QRPA calculations of $^{206}$Pb demonstrate that in the vicinity of the neutron threshold and above the theoretical dipole strength function fits a Lorentzian shape, as is generally 
assumed for the GDR \cite{Har01}. 
Thus, the combined analysis of the structure, transition densities, and energy-weighted sum rules of 
the 1$^-$ states provides a clean separation of the PDR and GDR excitations within the QRPA framework.
The QPM model basis is built of  QRPA phonons with spin and parity  
$J^\pi$=1$^{\pm}$,2$^+$, 3$^-$, 4$^+$, 5$^-$. The model Hamiltonian is  
diagonalized on an orthonormal set of wave functions constructed as a  
superposition of one-, two- and three-phonon components  
\cite{Tso04,Tso08}. The theoretical method allows for sufficiently  
large configuration spaces such that a unified description of  
one-phonon – PDR and GDR and multi-phonon states is feasible.
 Finally, from the individual transitions one may also compute experimental and theoretical \emph{cumulative} (or integrated) strengths below the neutron separation energy in order 
to visually assess the spectral contribution to the overall strength; see Fig.\,\ref{fig3}. As it is seen from the figure, the EDF+QPM calculations strongly suggest that the PDR dominates the distribution of $E1$ strength up to about $7$\,MeV, 
at which point the tail of the GDR starts making an important contribution. Overall, the PDR and the GDR account for about 77\% and  $12$\% of the $E1$ strength below the neutron separation energy, respectively. Below $8.1$\,MeV there is significant impact from multi-phonon states to the total $E1$ strength and to a lesser extent to the $M1$ strength;
$\approx$\,31\,$\%$ and $\approx$\,3\,$\%$, respectively. For these theoretical estimates, interference terms have been neglected but they are, of course, taken into account in the full calculations.
A detailed description of the elaborate multi-quasiparticle and multi-phonon approach 
has proven successful in various scenarios as well as additional improvements 
to the model for heavy nuclei displaying a high level density near threshold are beyond the scope 
of this letter. For a detailed discussion of the technical aspects of the model see 
Refs.\,\cite{Sol76,Tso04,Tso08,Ton10,Tso11,Sch13,Rus13,Pel15,Spi16}. 
The comparison of the theoretical results for low-energy electric and magnetic spectral distributions and cumulative strengths in $^{206}$Pb presented here with calculations given in Ref. \cite{End03} clearly reveals that the EDF+QPM approach provides a better description of the experimental data.

\begin{table*}
 \caption{Summary of a few moments of the photoabsorption cross section of $^{206}$Pb and $^{208}$Pb.}
 \begin{tabular}{lccccccr}
  \hline
  Nucleus & $E_{\rm max}$ & 60NZ/A  & $\sigma_{0}$ & $\sigma_{-1}$ & $\sigma_{-2}$ & Ref.\\
  & (MeV) & (mb\,MeV)  & (mb\,MeV) & (mb) & (mb/MeV) & \\
  \hline\rule{0pt}{2.5ex}
$^{206}$Pb &  26  & 2962 & 3544$\pm$294 & 241$\pm$17 & 18$\pm$1  & Present+\cite{Kon12, Har64}\\
  &   &  & 3437 & 240 & 18  & [ENDF] \\
  \\
  $^{208}$Pb &  25 & 2980 &  3981$\pm$331 & 287$\pm$18 &  20$\pm$1 & \cite{Vey70} \\
  &   &  & 3404 & 239 & 18  & [ENDF] \\
\hline
 \end{tabular}
 \label{table2}
\end{table*}

Having fully characterized the distribution of low-energy dipole strength in ${}^{206}$Pb, we now 
return to the main motivation behind this letter: what is the impact of this nuclear experiment on 
astrophysics? We start by assessing the impact of this measurement on the density dependence 
of the symmetry energy. Although the $E^{-2}$ weighting in Eq.\,(\ref{EDP}) makes the low-energy 
component of the dipole response of paramount importance to the evaluation of 
$\alphaD$ (or $\sigma_{-2}$), knowledge of the dipole response above the neutron separation 
energy is also required. Thus, to complement the present low-energy experiment we rely 
on the measurements of the photoabsorption cross section for $E_{\gamma}\!>\!S_{1n}$ reported in 
Refs.\,\cite{Kon12,Har64}. Using the combined experimental information, we list in Table\,\ref{table2}  a few moments of the experimental photoabsorption cross section. Approximately 90\% of the quoted uncertainties originate from the systematic uncertainties associated with the photon flux and detector efficiency measurements. 
As a reference, values for the classical TRK sum rule are
also provided. 
Beside these moments, additional predictions are 
displayed in Table\,\ref{table3} for the neutron-skin thickness of ${}^{206}$Pb (and ${}^{208}$Pb) together with a few bulk parameters of the symmetry energy; see Eq.\,(\ref{JLK}). These predictions were performed using a small set of accurately calibrated relativistic
EDFs\,\cite{Tod05,Che15}, with ground state properties computed in a mean-field approximation and the dipole response in a self-consistent RPA and the EDF+QPM\, (GiEDF) \cite{Tso04,Tso08} approaches; for a recent review on the imprint of the symmetry energy on the dipole response see Ref.\,\cite{Pie14}.  
Table\,\ref{table3} illustrates powerful correlations between nuclear observables. In the case of the
isovector dipole mode, the out of phase oscillation of protons against neutrons, the symmetry energy acts 
as the restoring force. In particular, theoretical models with a \emph{soft} symmetry energy, namely models 
that predict a slow increase of the symmetry energy with density (\emph{i.e., small} $L$) predict \emph{large} 
values for the symmetry energy at the sub-saturation densities of relevance to the excitation of this mode. 
In turn, this induces a quenching and hardening of the dipole strength relative to their stiffer counterparts
({\sl i.e.,} models with large $L$). 
Because the distribution of dipole strength is sensitive to the 
density dependence of the symmetry energy, the energy weighted sum (or total photoabsorption cross 
section $\sigma_{0}$) is not, as it is ``protected'' by the classical TRK sum rule. Indeed, 
Table\,\ref{table3} indicates a mild model dependence (of a few percent) in $\sigma_{0}$. Instead, 
the $E^{-2}$ weighting enhances the low-energy response and reveals a large sensitivity of
$\sigma_{-2}$ (of $\sim\!25\%$) to the density dependence of the symmetry energy. These findings 
suggest the following insightful correlation: the stiffer the symmetry energy, the larger the neutron-skin 
thickness of heavy nuclei, and the larger the electric dipole polarizability\,\cite{Rei10,Pie12,Pie14,Maz13}.
Based on the limited set of relativistic and nonrelativistic EDFs displayed in Table\,\ref{table3}, the measured experimental 
value of $\sigma_{-2}$ in ${}^{206}$Pb suggests a fairly soft symmetry energy, with values of its slope 
at saturation density in the range $48\!\lesssim\!L\lesssim\!60$\,MeV. Correspondingly, the extracted 
range of values of the neutron skin in ${}^{206}$Pb is $0.12\!\lesssim\!\Rskin^{206}\lesssim\!0.19$\,fm, 
which translates into $0.13\!\lesssim\!\Rskin^{208}\lesssim\!0.21$\,fm for ${}^{208}$Pb. 
From the EDF+QPM calculations of both, E1 and M1 strengths up to 25 MeV in $^{206}$Pb (the corresponding $\sigma_{0}$, $\sigma_{-1}$ and $\sigma_{-2}$ values are given in Table\,\ref{table3} with the notation GiEDF), we obtained for the total dipole polarizability $\alpha_D(^{206}$Pb)=127 mb/MeV=18.3 fm$^3$/e$^2$. The comparison of the $\alpha_D$ value with the one obtained in ${}^{208}$Pb \cite{Tam11} shows a decrease of ~3$\%$ in ${}^{206}$Pb. This is found correlated with the decrease of the neutron skin thickness of $^{206}$Pb, which is ~4$\%$ less than that of ${}^{208}$Pb (see in Table\,\ref{table3}).

 \begin{table*}
 \caption{Moments of the photoabsorption cross section of $^{206}$Pb as in Table\,\ref{table2},
 but now as predicted by a series of accurately calibrated relativistic EDFs\,\cite{Tod05,Che15} and the non-relativistic EDF (GiEDF)  underlying the QPM approach.
 Also shown is the neutron-skin thickness of $^{206}$Pb (and $^{208}$Pb displayed in square
 brackets) as well as values of the symmetry energy ($J$) its slope ($L$) and its curvature
 ($K_{\rm sym}$) at saturation density [see Eq.(\ref{JLK})]. The large negative GiEDF value for $K_{\rm sym}$ is typical for non-relativistic approaches, see e.g. \cite{Ina15}. }
 \begin{tabular}{lccccccc}
 \hline
  Model & $\sigma_{0}$ & $\sigma_{-1}$ & $\sigma_{-2}$ & $\Rskin$ & $J$ & $L$  & $K_{\rm sym}$\\
             &   (mb\,MeV)   &      (mb)           &     (mb/MeV)    &       (fm)       & (MeV) & (MeV) & (MeV) \\
  \hline\rule{0pt}{2.5ex}
 RMF012     & 3653 & 237 & 17 & 0.12 [0.13] & 29.8 & 48.3 & $\phantom{-}98.7$ \\
 FSUGarnet & 3689 & 243 & 18 & 0.15 [0.16] & 30.9 & 51.0 & $\phantom{-}59.5$ \\
 FSUGold    & 3638 & 251 & 19 & 0.19 [0.21] & 32.6 &  60.5 & $-51.3$ \\
 RMF028     & 3711 & 265 & 21 & 0.26 [0.29] & 37.5 & 112.6 & $\phantom{-}26.2$ \\
 RMF032     & 3812 & 262 & 21 & 0.30 [0.32] & 41.3 & 125.6 & $\phantom{-}28.6$ \\
 GiEDF        &  3060& 230 & 18& 0.15 [0.16]& 33.4& 53.9 & -188.4\\
 \hline
 \end{tabular}
 \label{table3}
\end{table*}

\begin{figure}
\centering
\includegraphics[width=8.5cm,clip]{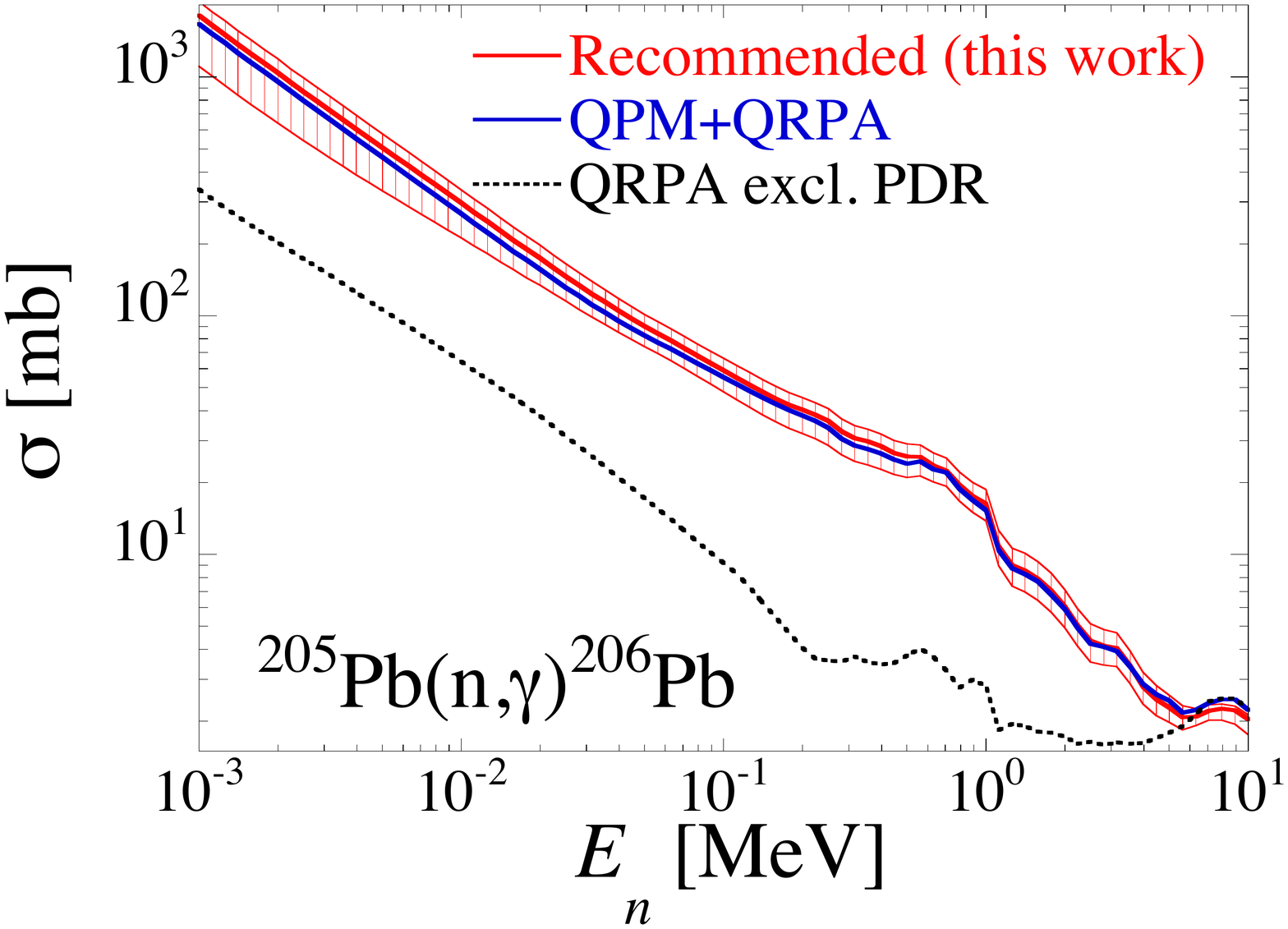}
\caption{(Color online) 
Radiative capture cross section $^{205}$Pb(n,$\gamma$)$^{206}$Pb 
using as input the experimental $E1$ and $M1$ dipole strength 
(red curve) or the three-phonon EDF+QPM plus EDF+QRPA predictions (blue curve). 
The dotted line is obtained with the EDF+QRPA strength excluding the PDR contribution.} 
\label{fig4}
\end{figure}

We now proceed to assess the impact of the present measurement on the neutron capture cross section, 
and possibly on stellar nucleosynthesis. The intensity and the energy distribution of the nuclear 
dipole response,
including both the low-energy PDR plus the contributions from core polarization 
below the neutron separation energy,
are fundamental ingredients for the determination of the 
neutron-capture rates \cite{Rau13, Tso15,Gor98}. To estimate the radiative neutron cross section of $^{205}$Pb, 
statistical model calculations using the TALYS-1.8 code\,\cite{Koning12} have been carried out, 
with the results displayed  in Fig.\,\ref{fig4}. The ``Recommended'' curve corresponds to the radiative 
cross section obtained with the presently measured $E1$ and $M1$ $\gamma$-ray strength, 
complemented by EDF+QPM predictions outside of the experimental energy range. The uncertainty 
band in Fig.\,\ref{fig4} stems from the experimental uncertainties associated with the $\gamma$-ray strength, but is also due to the use of different models to predict the nuclear level density\,\cite{Koning08,Capote09}.
The recommended cross section is obtained with the combinatorial model of Ref.\,\cite{Goriely08}. 
Note that level densities are constrained on the cumulative number of low-lying levels, including the 
$J\!=\!1$ states measured in the present experiment. As shown in Fig.\,\ref{fig4}, the QPM  
model supplemented with QRPA calculations generate a cross section that is in excellent agreement 
with the calculations based on the experimental strength. Whereas the $M1$ contribution is found to be rather insignificant (less than 5\%), the combined PDR plus 
core polarization contribution is crucial for a proper description of the cross section. Indeed, excluding the
PDR contribution, QRPA predictions by themselves yield negligible $E1$ strength below 6\,MeV, leading 
to a cross section about 5 times lower relative to the one involving the combined contribution, as shown in 
Fig.\,\ref{fig4}. Note that the contribution of higher multipolarities has also been included in the calculation of the radiative neutron capture cross section but remains negligible. Based on these results, the experimentally constrained Maxwellian-averaged cross section at 30\,keV is estimated 
to be $130\!\pm\!25$\,mb---a value that is consistent with the prediction of $125\!\pm\!22$\,mb estimated 
solely by theoretical means. This latter value has been recommended in Ref.\,\cite{Bao00} and has been 
traditionally used in s-process calculations. With this updated cross section, the dynamics involved in the
survival and destruction of $^{205}$Pb in AGB stars is put on much more solid ground.

In conclusion, experimental high-resolution studies and EDF+QPM predictions of the electromagnetic response of  $^{206}$Pb permit a separation of the PDR from the tail of the GDR and multi-phonon excitations due to core polarization effects. Our findings suggest that the low-energy dipole strength is predominantly 
electric in character and mainly due to a PDR skin oscillation. However, a substantial contribution from both
the low-energy tail of the GDR and multi-phonon states to the total $E1$ strength is also observed that is
responsible for the fragmentation pattern of low-energy dipole states. Moreover, the EDF+QPM theory 
successfully reproduces the low-energy $M1$ spectral distribution, suggesting that it is mostly due to 
spin-flip excitations. In combination with relativistic EDFs that are accurately calibrated to 
ground-state properties of finite nuclei, estimates for the neutron-skin thickness in both $^{206}$Pb 
and $^{208}$Pb are provided, with the latter consistent with some recent 
analyses\,\cite{Tam11}. In turn, these estimates suggest a relatively soft symmetry energy. In the 
context of stellar nucleosynthesis, an updated---experimentally constrained---Maxwellian-averaged 
radiative capture cross section for $^{205}$Pb(n,$\gamma$)$^{206}$Pb is obtained. The work reported 
here illustrates the vital and ever increasing role that measurements of exotic modes of excitation in 
neutron-rich nuclei are playing in the determination of observables of critical astrophysical importance. 

This work was performed under the auspices of US DOE by LLNL under contract DE-AC52-07NA27344 with partial support from the LDRD Program under project 16-ERD-022
and based upon work supported by the US DOE Office of Science, Office of Nuclear 
Physics under Award Numbers DE-FG02-97ER41033, DE-FG02-97ER41041, DE-FG02-97ER41042, 
DE-FG52-06NA26155, and DE-FD05-92ER40750, as well as DFG grant Le439/6 and the F.R.S.-FNRS.

\vfill\eject

\end{document}